\documentclass[aps,pra,preprint,nofootinbib,showkeys,%
showpacs,superscriptaddress]{revtex4}

\usepackage{graphicx}
\usepackage{bm}
\usepackage{amsmath}

\newcommand{\im}{\text{Im}\,}
\begin{document}
\title{Lifshitz formula by spectral summation method}
\author{V.~V.~Nesterenko}
\email{nestr@theor.jinr.ru} \affiliation{Bogoliubov Laboratory
of Theoretical Physics, Joint Institute for Nuclear Research,
Dubna 141 980, Russia}
\author{I.~G.~Pirozhenko}
\email{pirozhen@theor.jinr.ru} \affiliation{Bogoliubov
Laboratory of Theoretical Physics, Joint Institute for Nuclear
Research, Dubna 141 980, Russia} \affiliation{Dubna
International University, Dubna 141 980, Russia}

\date{\today}
\begin{abstract}
The Lifshitz formula is derived by making use  of the
spectral summation method which  is a mathematically
rigorous simultaneous application of both the mode-by-mode
summation technique and scattering formalism. The
contributions to the Casimir energy of electromagnetic
excitations of different types (surface modes, waveguide
modes, and photonic modes) are clearly retraced. A correct
transition to imaginary frequencies is accomplished with
allowance for all the peculiarities of the frequency
equations and pertinent scattering data in the complex
$\omega $ plane, including, in particular, the cuts
connecting the branch points and complex roots of the
frequency equations (quasi-normal modes). The principal
novelty of our approach is a special choice of appropriate
passes in the contour integrals which are used for
transition to imaginary frequencies. As a result, the long
standing problem of cuts in the complex $\omega $ plane is
solved completely. Some subtleties and vague points in
previous derivations of the Lifshitz formula are
elucidated. For completeness of the  presentation, the necessary
mathematical facts are  also stated, namely, solution of the
Maxwell equations for configurations under consideration,
scattering formalism for parallel plane interfaces,
determination of the frequency equation roots, and others.
\end{abstract}
\maketitle
\section{\label{intr}Introduction}
It is not  overstatement to say that the Lifshitz formula
is the basis for practically all the Casimir calculations
dealing with plane boundaries \cite{book}. However, it
was recognized  in the literature
\cite{Spruch-1,Langbein-SSC,Langbein-JChPhys,Sernelius} that the
derivation of this formula in the original paper
\cite{Lif-DAN,Lif} is very complicated. It proved that
especially involved is the transition to the imaginary
frequencies when  obtaining the Lifshitz formula by
deforming the contours of integration in the complex
frequency plane. As far as we know, nobody has
succeeded in repeating this part of Lifshitz' calculations.
Furthermore, the Lifshitz paper \cite{Lif} did not trace in detail the
contributions to the vacuum energy  generated by different
branches of the electromagnetic excitation spectrum
(propagating waves and evanescent waves) and, in
particular, different presentations of these contributions
in terms of real and imaginary frequencies.
Now these  points have become important in searching for the materials
with the properties necessary for obtaining  the desired
characteristics of the Casimir forces.

Later on, the Lifshitz formula was obtained by making use
of different methods, namely: the quantum field theory
technique in condensed matter physics \cite{DLP1,DLP2},
mode-by-mode summation \cite{book,KNSchram,Schram,Schram-thesis,%
NPerW,Gerlach,Langbein-SSC,Langbein-JChPhys,KMM}, methods
of quantized surface modes and quantized Fresnel modes
\cite{Spruch-2,Spruch-3,Spruch-4}, scattering formalism
\cite{scatt-L,scatt-L1,scatt-L2}, local Green's functions
\cite{LL,Milton-book,Brevik}, and so on.
These approaches are quite different, and at  first
glance  it is difficult to reveal their interrelation
although these methods lead to the same Lifshitz formula,
which has never been in doubt.

For the sake of understanding this situation, we propose in
the present paper  another derivation of the Lifshitz
formula, namely, this formula will be obtained by making
use of the spectral summation method. This approach is  a
mathematically rigorous simultaneous application of both
the mode-by-mode summation technique and the scattering
formalism. At the same time, some subtleties and vague
points in previous derivations of the Lifshitz formula will
be elucidated.

Our derivation of the Lifshitz formula includes two steps:
First, we determine the spectrum of electromagnetic
excitations for given boundaries, taking into account the
material properties of the media under consideration
\cite{plasma-shell}, and afterward we accomplish the summation
of the  relevant zero-point energies
\cite{Milton-book,Milton-R,RNC} connected with all branches
of this spectrum. For a description of the electromagnetic
fields in matter, we shall use the dielectric formalism with
allowance for the usual temporal dispersion. In the
framework of this approach, one can concentrate
on the  electromagnetic field dynamics, while the dynamics
of induced charges and currents is taken into account by
introducing the permittivity and permeability
$\varepsilon(\omega),\; \mu(\omega)$.\footnote{The
electromagnetic field is coupled, through the Maxwell
equations, with charges and currents, therefore one can
take, as a dynamic variable, the local displacement of
a continuous charged  liquid that describes the free
electrons inside the medium \cite{Romania}.} Some
nontrivial points in calculating the electromagnetic energy with
allowance for the media dispersion in the Casimir studies
are recently  clarified in Ref.\ \cite{Milton-dis}.

When determining the spectrum of the electromagnetic field, all
physically relevant solutions to the Maxwell equations should
be taken into account. Obviously, such solutions are all
squared integrable solutions which belong to the $L_2$
functional space. In addition, this set of solutions should
be complete. In the rigorous spectral theory of differential
operators \cite{Newton-book,Yafaev}, the following
assertion is proved: the functional space including the
solutions that describe the bound states (discrete
spectrum branch) and scattering states (continuous branch
of the spectrum) is complete with respect to the
$L_2$ norm. It is these solutions that we shall find in the
problem under study and on this basis we shall determine
the relevant spectrum, i.\ e., the admissible
values of the electromagnetic oscillation frequencies
$\omega $.\footnote{In the general case, the spectrum of
the differential operator may be more complicated
\cite{Naimark}. However, in the Casimir calculations such
problems do not occur.} Upon determining the complete set
of solutions to the Maxwell equations, quantization of
the electromagnetic field is trivial because we
deal here with the sum of two infinite sets (discrete
and continuous) of noninteracting oscillators. The discrete
set of oscillators describes the evanescent waves including
the surface and waveguide modes,\footnote{In physical
problems dealing with the plane boundaries,  the surface
modes and waveguide modes are called sometimes  the
guided modes~\cite{gm}.} and a continuous set of oscillators
is responsible for the scattering states. All these modes
will be rigorously defined in the course of constructing
the complete set of solutions to the Maxwell equations
(see below).

The summation of  zero-point energies of the discrete modes
can be accomplished, obviously, in a straightforward way,
while for the summation of such energies appertaining to
scattering states, one has to take advantage of the
scattering formalism \cite{NVV-bc}. Preceding derivations
of the Lifshitz formula in the framework of the
mode-by-mode summation alone or by making use of the scattering
formalism only seem somewhat contradictory at first glance. Indeed,
each of
these methods can be rigorously applied  to one branch of
the spectrum separately, namely, the mode-by-mode summation is
applicable to the discrete part of the spectrum solely and the
scattering formalism enables one to take into account the
scattering states, i.\ e.\ it is applicable to the
continuous part of the electromagnetic spectrum only.

However, it turns out that the ``naive'' presentation, by the contour
integral in the complex frequency plane, of the
contribution to the vacuum energy generated by one spectrum
branch automatically  incorporates the contribution of the
other branch. In order to show this, we perform rigorous
summation of the zero-point energies appertaining to
both the  spectrum branches (discrete and continuous ones). It
is this approach that enables one to overcome known
drawbacks in this field, for example, to take into account
the cuts in the complex frequency plane
\cite{Langbein-SSC,Langbein-JChPhys,NPerW,Gerlach,Schram,Schram-thesis},
to define correctly the contributions to the vacuum energy
due to different branches of the electromagnetic spectrum \cite{Gerlach,Schram},
and so on.

The important step in the Casimir calculations is the
transition to the imaginary frequencies. In our opinion
this transition was not rigourously justified at least in
the framework of the mode-by-mode summation technique. It
is this issue that resulted in some confusions when deriving   the
Lifshitz formula by the mode summation
\cite{KNSchram,Schram,Schram-thesis,%
NPerW,Gerlach,Langbein-SSC,Langbein-JChPhys,%
Spruch-2,Spruch-3,Spruch-4} (see Conclusion in the present paper).
In order to accomplish  this transition in a
consistent way, the analysis of the analytical properties of
the frequency equations and the scattering data in the
complex $\omega$ plane should be conducted as a preliminary, and
the needed cuts connecting the branch points in this plane
should be done. In the pertinent literature
\cite{book,Spruch-1,Langbein-SSC,Langbein-JChPhys,%
KNSchram,Schram,Schram-thesis,NPerW,KMM,%
Gerlach,Spruch-2,Spruch-3,Spruch-4,%
scatt-L,scatt-L1,scatt-L2,LL} these issues were not  examined.
We are going to fill in this gap.

It is worth noting here that the analytic properties of the
scattering matrix (or the Jost function) in the Casimir
studies are different in comparison with the standard
theory of potential scattering. In fact, they are close to
those for the Klein-Gordon equation, the role of mass
squared being played by $k^2$, where $k$ is the wave vector
along the unbounded dimensions. This implies, in
particular, that the analytic properties of the
scattering matrix with respect to the complex frequency
$\omega$ in the Casimir calculations should be revealed by
direct analysis of its explicit form without referring to
the nonrelativistic potential scattering in the framework
of the Schr\"odinger equation.

The layout of the paper is the following. Section
\ref{spectr} is devoted to the spectrum  of electromagnetic
excitations in the problem under consideration. For the
Casimir studies the materials permitting the surface waves are of
current interest. The permittivity of these media can
acquire, in certain frequency bands, negative values.
Typical examples of these materials (metals and isotropic
dielectrics) are discussed briefly. In
Sec.\ \ref{Maxwell}, the general solutions to the Maxwell
equations for plane parallel interfaces are constructed.
Both branches of the spectrum are considered in detail.
In Sec.\ \ref{Casimir} the Lifshitz formula for the Casimir
energy is derived by summing up the zero-point energies of
discrete modes and scattering states. In Sec.\ \ref{Conclusion}
(Conclusion), the obtained results are summed  and unsolved
problems in this field are  briefly outlined.

\section{\label{spectr}The spectrum of electromagnetic excitations}
\label{spectr}
The electromagnetic excitations in the media are diverse,
and they essentially depend on the dielectric and magnetic
properties of the background. For simplicity, only
isotropic media  are considered in this paper.

In the Casimir studies we are interested in the long-wave
excitations with  $\lambda \gtrsim d $, where $d$ is
the characteristic scale in this field ($d
\sim $ 10 -- 100 nm), and $\lambda$ is the considered wave
length. The description of the condensed matter in terms of
the dielectric permittivity assumes the long wave
approximation as well.

In applications the dielectric properties of the materials
play the leading role; therefore in what follows we put the
magnetic permeability equal to one.

Of particular interest for theoretical and experimental
Casimir studies are the media with dielectric permittivity
taking on negative values over a certain frequency range.
Only interfaces of such media support surface
electromagnetic waves which can be treated as collective
excitations of electron density. The surface waves
together with waveguide solutions belong to the discrete
branch of the spectrum (see Sec.\  \ref{branches}).

Metals are typical media of this sort. The overall picture
of the electromagnetic waves  (excitations) in metal
resembles that in plasma but is not identical with it. In the
bulk of the metal  the local electron density oscillates
with the characteristic plasma frequency
\begin{equation}
\label{omega-pl} \omega_{p}^2=\frac{4 \pi  n e^2}{m}\,{,}
\end{equation}
where $n$ is the mean electron density in the metal and $m$
is the electron mass, the role of the restoring force being
played by the electrostatic field. The volume plasma
excitations give rise to surface excitations of the
electron density (surface plasmons) which propagate along
the metal-dielectric or metal-vacuum interface. Obviously,
the oscillations of the free charge  density cause,
according to the Maxwell equations, the oscillations of the
electromagnetic field.

For metals the following dielectric function is commonly used (plasma model)
\begin{equation}
\label{m} \varepsilon(\omega)=1-
\frac{\omega_{p}^2}{\omega^2}\,{,}
\end{equation}
where $\omega_{p}$ is the plasma frequency
(\ref{omega-pl}). The dielectric permittivity
$\varepsilon(\omega)$ takes negative values in the region
$\omega < \omega_{p}$. The field oscillations brought about  by the
surface excitations of the electron density are also
localized near the surface and  are
referred  to as the  surface plasmons (see the corresponding analysis of
the solutions to the Maxwell equations in Sec.\
\ref{Maxwell}).

Inside  the metal, the bulk waves propagate with the frequencies
$\omega>\omega_{p}$. The transversal and longitudinal bulk
waves form the continuous branch of the spectrum.


It is worth noting here the following. In order to describe
the dielectric  properties of a metal more precisely,
instead of the simple plasma model (\ref{m}), one has to use the
Drude model \cite{book}, which takes into account dissipation,
\begin{equation}
\label{Drude} \varepsilon_{D}(\omega)=1-
\frac{\omega_{p}^2}{\omega(\omega+i\gamma)}\,{,}
\end{equation}
where $\gamma$ is the relaxation parameter. However, all
this immediately results  in the necessity to consider the
complex-valued eigenfrequencies. So far, the spectral
summation method has not been extended to this case (see also
Sec.\ \ref{Conclusion} Conclusion). Therefore,  we shall further
describe the dielectric function for metal by the simple plasma
model\footnote{At separations below 1~$\mu$m, the Drude and
plasma models in the Lifshitz formula lead to the values of the
Casimir energy and Casimir force differing by less than
2\%. One might expect that at such separations thermal
corrections are not important~\cite{book}.} (\ref{m}).

In the isotropic dielectric the permittivity is described with
a good accuracy by an oscillator model \cite{Born-Wolf}. To
gain better understanding, one can consider the atoms in
a dielectric as weakly damped harmonic oscillators with
eigenfrequencies  $\omega_i$ excited by an external
electric field $\mathbf{E}=\mathbf{E_0} e^{-i\omega t}$. The
$N$-oscillator model gives the following dielectric function:
\begin{equation}
\label{d} \varepsilon(\omega)= \varepsilon(\infty)+\sum_{j}^N\frac{S_j \,\omega^2_j}{\omega^2_j
-\omega^2-i \Gamma_j \omega},
\end{equation}
where $S_j$ is the $j$th oscillator strength  and $\Gamma_j$ is
its relaxation parameter. In what follows, we neglect again
the absorption (undamped oscillators, $\Gamma_j=0$) and
consider real dielectric permittivity. It is negative for
$\omega$ approaching $\omega_j$ from the right.

Thus,  the condition for the existence of the surface plasmon at
the flat  dielectric-vacuum interface  is fulfilled near the
narrow absorption lines of the dielectric media. The number
of the plasmons in this case is equal to the number of
oscillators in the model.

The bulk waves in dielectrics (continuous spectrum) lie in
the ranges of frequencies where $\varepsilon(\omega)$ is
positive.

\subsection{\label{Maxwell}General solutions to the Maxwell equations}
First,  we recall briefly the formulation of  the Maxwell theory
for compound media with constant permittivity $\varepsilon$
and permeability  $\mu $ in each region. In this case, the
harmonic in time  electric ($\mathbf{E}$) and magnetic
$(\mathbf{H})$ fields are described by the Maxwell
equations~\cite{Stratton}
\begin{eqnarray}
\bm{\nabla}\times
\mathbf{E}&=&i\,\frac{\omega}{c}\,\mathbf{B}\,{,}\qquad
\bm{\nabla}\cdot \mathbf{D}=0\,{,} \label{a-1}\\
 \bm{\nabla}\times
\mathbf{H}&=&-i\,\frac{\omega}{c}\, \mathbf{D},\qquad
\bm{\nabla}\cdot \mathbf{B}=0\,{,} \label{a-2}
\\\mathbf{D}&=&\varepsilon \mathbf{E}\,{,}\qquad \mathbf{B}=\mu
\mathbf{H}\,{,}\quad \mathbf{x}\notin \Sigma\,{,}\label{a-3}
\end{eqnarray}
which hold outside the interface $\Sigma$ separating the
regions with different $\varepsilon $ and $\mu $, and by
matching conditions on $\Sigma$. These conditions require the
continuity of tangential components of the fields $\mathbf{E}$ and
$\mathbf{H}$ when crossing $\Sigma$:
\begin{equation}
\label{a-4} \mbox{discont } (\mathbf{E}_{\parallel}) =0,\quad
\mbox{discont } (\mathbf{H}_{\parallel})=0 \,{.}
\end{equation}

The Gauss units are used and it is assumed that external charges
and currents are absent (both volume and surface ones). The
common time factor $e^{-i\omega t}$ will be dropped.

When allowing for the time dispersion, the
permittivity $\varepsilon$  and permeability $\mu
$ in material equations  (\ref{a-3}) and further
should be treated as functions of the  frequency $\omega$.

General solution to Eqs.\ (\ref{a-1}), (\ref{a-2}), and (\ref{a-3}) can be
represented in terms of two independent Hertz vectors in the
following way~\cite{Stratton,HdP1,HdP2}:
\begin{eqnarray}
\mathbf{E}=\bm{\nabla}\times\bm{\nabla}\times\bm{\Pi}', &\quad&
\mathbf{H}=-i\varepsilon\,\frac{\omega}{c}\,\bm{\nabla}\times\bm{\Pi}'\quad (\text{TM-modes});\label{a-5}\\
\mathbf{E}=i\mu
\,\frac{\omega}{c}\,\bm{\nabla}\times\bm{\Pi}'',&\quad&
\mathbf{H}=\bm{\nabla}\times\bm{\nabla}\times\bm{\Pi}'' \quad
(\text{TE-modes})\,{.} \label{a-6}
\end{eqnarray}
Here $\mathbf{\Pi '}$ is the electric Hertz vector, $\mathbf{\Pi
''}$ is  the  magnetic Hertz vector, and $c$ is the velocity of
light in vacuum.

In each region with given $\varepsilon(\omega)$ and
$\mu(\omega)$ the Hertz vectors obey the Helmholtz vector
equation
\begin{equation}
\left ( \bm {\nabla}^2 + k^2\right )\bm{\Pi}=0\,{,} \label{a-7}
\end{equation}
where the wave number $k$ is given by
\begin{equation}
\label{a-8} k^2=\varepsilon (\omega)\mu (\omega)\,\frac{\omega^2}{c^2}\,{.}
\end{equation}

The boundary conditions (\ref{a-4}) involve the complete fields
$\mathbf{E}$ and $\mathbf{H}$, i.e.,  the sums of the TE and
TM modes. Fortunately, for some interface geometries these
conditions  do not couple the TE and TM polarizations. It
is true for plane interfaces (see below), for  spheres,
and in some other cases \cite{book,Stratton,Milton-book,Milton-R,RNC}.

It is  known that in the source-free case the general
solution of Maxwell's equations can be derived from two
real scalar functions~\cite{Whittaker,GW,Nisbet} which may
be chosen in different ways. Thus, the Hertz potentials have
in fact only one nonvanishing component. The precise choice
of these components is determined by the interface geometry
and for reasons of simplicity and convenience.

We consider the flat interface parallel to  the $(x,y)$ coordinate plane and cutting
the $z$ axis at the point $z=z_0$. Let $\mathbf{e}_x,\;\mathbf{e}
_y,\;\mathbf{e}_z$ be the unit base vectors in the chosen coordinate
system. The $z$ components  of the Hertz potentials are treated
as two independent functions in the general solution to the
Maxwell equations:
\begin{equation}
\label{a-9} \bm{\Pi}'= \mathbf{e}_z e^{i\mathbf{k\cdot
s}}\Phi(z),\quad \bm{\Pi}''= \mathbf{e}_z e^{i\mathbf{k\cdot
s}}\Psi(z)\,{.}
\end{equation}
Here $\mathbf{k}$ is a two-component wave vector parallel to the
interface, $\mathbf{k}=(k_x,\,k_y)$, and $\mathbf{s}=(x,y)$. The
common time-dependent factor $e^{-i\omega t}$ is dropped as
usual.

By substituting the Hertz potentials (\ref{a-9}) in (\ref{a-5}) and
(\ref{a-6}) we get the fields for the TE modes,
\begin{eqnarray}
\mathbf{E}&=&\mu\,\frac{\omega}{c}\,k_x\,\mathbf{e}_y\,
e^{i\mathbf{k\cdot s}}\Psi(z),
\label{a-10}\\
\mathbf{H}&=&ik_x \mathbf{e}_x\,e^{i\mathbf{k\cdot s}}\Psi'(z)+\mathbf{e}_z
k^2e^{i\mathbf{k\cdot s}} \Psi(z)\label{a-11}
\end{eqnarray}
and for the TM modes,
\begin{eqnarray}
 \mathbf{E}&=&ik_x\mathbf{e}_x e^{i\mathbf{ks}}\Phi'(z)+\mathbf{e}_z
k^2e^{i\mathbf{ks}} \Phi(z)\label{a-12} \\
\mathbf{H}&=&-\varepsilon\,\frac{\omega}{c}\, k_x\mathbf{e}_y e^{i\mathbf{k\cdot
s}}\Phi(z)\,{.}
\label{a-13}
\end{eqnarray}
Without loss of generality, we directed the $x$ axis along
the vector $\mathbf{k}$: $\mathbf{k}=(k_x=\pm k, 0)$.

For the fields (\ref{a-10})--(\ref{a-13}) to obey the Maxwell
equations (\ref{a-1})--(\ref{a-3}) the Hertz potentials should
meet the Helmholtz equation (\ref{a-7}), which now assumes  the
form
\begin{eqnarray}
\label{a-14} -\Phi''(z)&=&\left
(\varepsilon\mu\,\frac{\omega^2}{c^2}\,-k^2 \right ) \Phi(z)
\quad \text{(TM modes)}\,{,}
\\
-\Psi''(z)&=&\left (\varepsilon\mu\,\frac{\omega^2}{c^2}\,-k^2 \right )
\Psi(z)\quad\text{(TE modes)}\,{,} \label{a-15}\\
-\infty <z<\infty,&& z\neq z_0, \quad  k^2=k_x^2+k_y^2\,{.}
\nonumber
\end{eqnarray}

Substitution of the fields [\ref{a-10})--(\ref{a-13}] into the
continuity equations (\ref{a-4}) results in the matching
conditions at $z=z_0$ separately  for the functions $\Phi$ and $\Psi$
\begin{eqnarray}
&\left [ \varepsilon(z_0)\Phi(z_0) \right]=0, \quad \left [ \Phi'(z_0)
\right]=0&\quad \text{(TM modes)}\,{,}\label{a-16}
\\
&\left [\mu(z_0) \Psi(z_0) \right]=0, \quad \left [ \Psi'(z_0)
\right]=0&\quad \text{(TE modes)}\,{,}\label{a-17}
\end{eqnarray}
where the notation
\begin{equation}
\label{a-18} \left [ F(z) \right]\equiv F(z+0)-F(z-0)
\end{equation}
is introduced.

For a given value of $k^2>0$ the differential equations
(\ref{a-14}) and (\ref{a-15}), with matching conditions
(\ref{a-16}) and (\ref{a-17}), and physical conditions at
infinity  $z\to \pm \infty$ result in two spectral problems
for the TE and TM polarizations. The frequency $\omega$ plays
part of the spectral parameter. For complicated functions
$\varepsilon(\omega)$ and $\mu(\omega)$  the spectral
parameter may enter into the initial   differential equations
nonlinearly.

\subsection{\label{branches}The branches of electromagnetic spectrum
relevant for calculation of vacuum energy}

In order to calculate the vacuum energy of the  electromagnetic field,
one has preliminarily to quantize this field with allowance for given boundary
conditions. To this end,
the full set of solutions to the Maxwell equations
including all physically relevant ones is needed. In the
directions parallel to the interfaces the electromagnetic
field dynamics is free. Therefore, we are  concerned with
the behavior of the solutions normal to the boundaries,
along the $z$ axis.

As was mentioned in the Introduction, the full set of
physically relevant solutions in the problem under
consideration comprises discrete natural modes and
scattering states.

Normal modes are the solutions to the Maxwell equations
which are localized near and between the interfaces, their
energy being localized too. They are analogous to the
bound states in quantum mechanics. These solutions are
square-integrable on the whole $z$ axis. The corresponding
frequencies (the eigenvalues of the spectral problem) take
on discrete real values.

The scattering states (propagating modes) are described by
incident waves  and outgoing scattered (or reflected and
transmitted) waves. Their frequencies are real and positive
forming a continuous spectrum.

\subsubsection{\label{Discrete}Discrete part of the spectrum}
We start with considering a single flat interface in order
to elucidate the properties of dielectric and magnetic
functions required for the existence of normal modes in the
problem under study.

\paragraph{Normal modes for a single plane interface.}
Let the plane $z=0$ separate two uniform half spaces
characterized by $\varepsilon_1,\,\mu_1$ and
$\varepsilon_2,\,\mu_2$, or symbolically
$(\varepsilon_1,\mu_1|\varepsilon_2,\mu_2)$. The general
solutions to the Maxwell equations are defined by two
functions $\Phi(z)$ and $\Psi(z)$ obeying  (\ref{a-14}),
(\ref{a-15}) and the matching conditions (\ref{a-16}),
(\ref{a-17}).  The eigenfunctions in the problem should
decrease exponentially in both directions from the
interface, i.e., they behave like outgoing waves with
imaginary wave vectors,

\begin{equation}
\label{2-4} \Phi(z) =\begin{cases} A_1e^{-ik_1z},&  z<0, \cr
A_2e^{ik_2z},& z>0\,{,}\cr
\end{cases}
\end{equation}
where $ A_1$ and $A_2$ are the constant amplitudes,
\begin{equation}
\label{2-5} k_\alpha^2=\varepsilon_\alpha
\mu_\alpha\frac{\omega^2}{c^2}-k^2=\frac{\omega^2}{c^2_\alpha}-k^2<0, \quad
c_\alpha^2=\frac{c^2}{\varepsilon_\alpha\mu_\alpha},\quad
\alpha=1,2\,{,}
\end{equation}
provided that
\begin{equation}
\label{2-6} k_\alpha=+i\mid k_\alpha \mid \,{,} \quad
\alpha=1,2\,{.}
\end{equation}
On substituting (\ref{2-4}) and the analogous representation for
the function $\Psi (z)$ into the matching conditions
(\ref{a-16}) and (\ref{a-17}) we arrive at the following
equations determining the eigenfrequencies
\begin{eqnarray}
  \varepsilon_1 k_2+\varepsilon_2 k_1 &=&0 \quad \text{(TM-modes),} \label{2-7} \\
  \mu_1 k_2+\mu_2 k_1 &=&0 \quad \text{(TE-modes).} \label{2-8}
\end{eqnarray}
In view of condition (\ref{2-6}) the frequency equations
(\ref{2-7}) and (\ref{2-8}) may have real roots (real
eigenfrequencies) only if the  permittivities
$\varepsilon_1,\varepsilon_2 $ and  permeabilities $\mu_1, \mu_2
$ have different signs.

\begin{figure}[p!]
\noindent\centerline{
\begin{tabular}{lr}
\parbox{84.0mm}{\includegraphics[width=84.0mm]{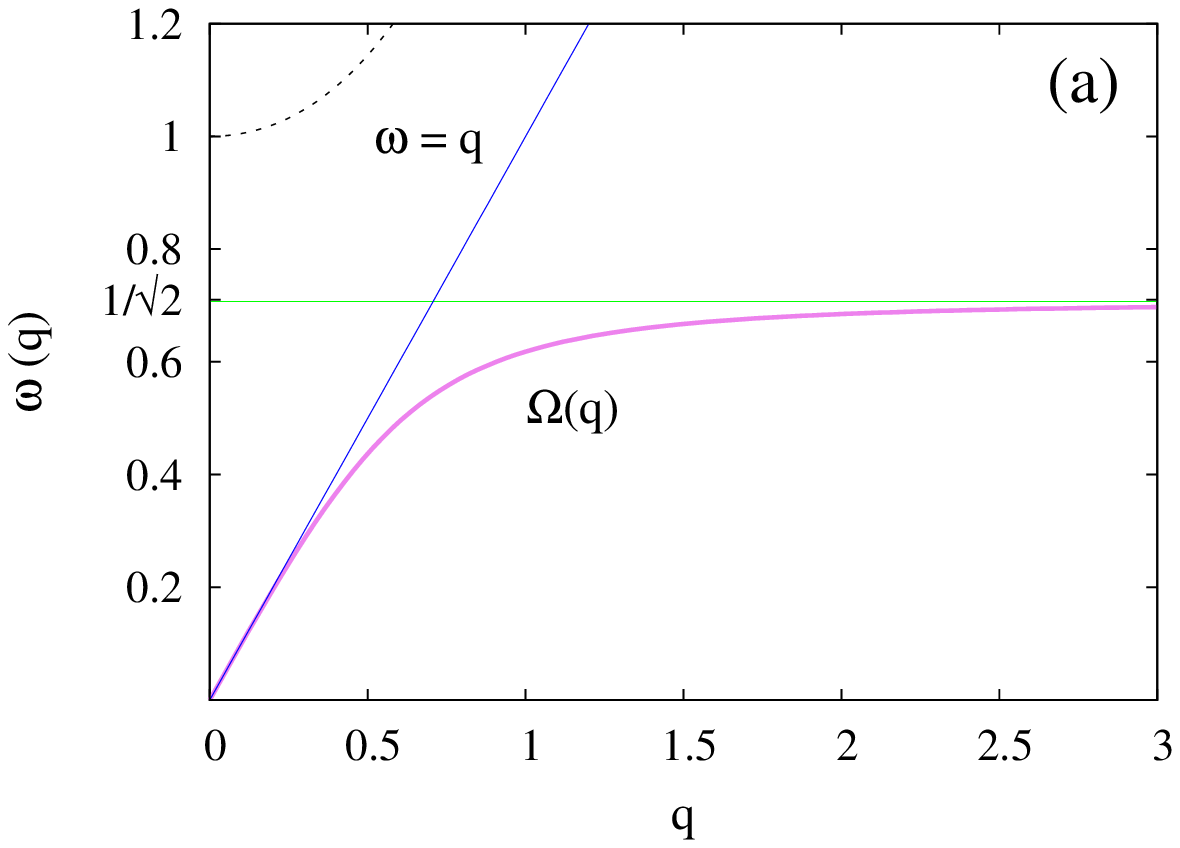}\hfill\vskip4mm}&
\parbox{84.0mm}{\hfill\includegraphics[width=84.0mm]{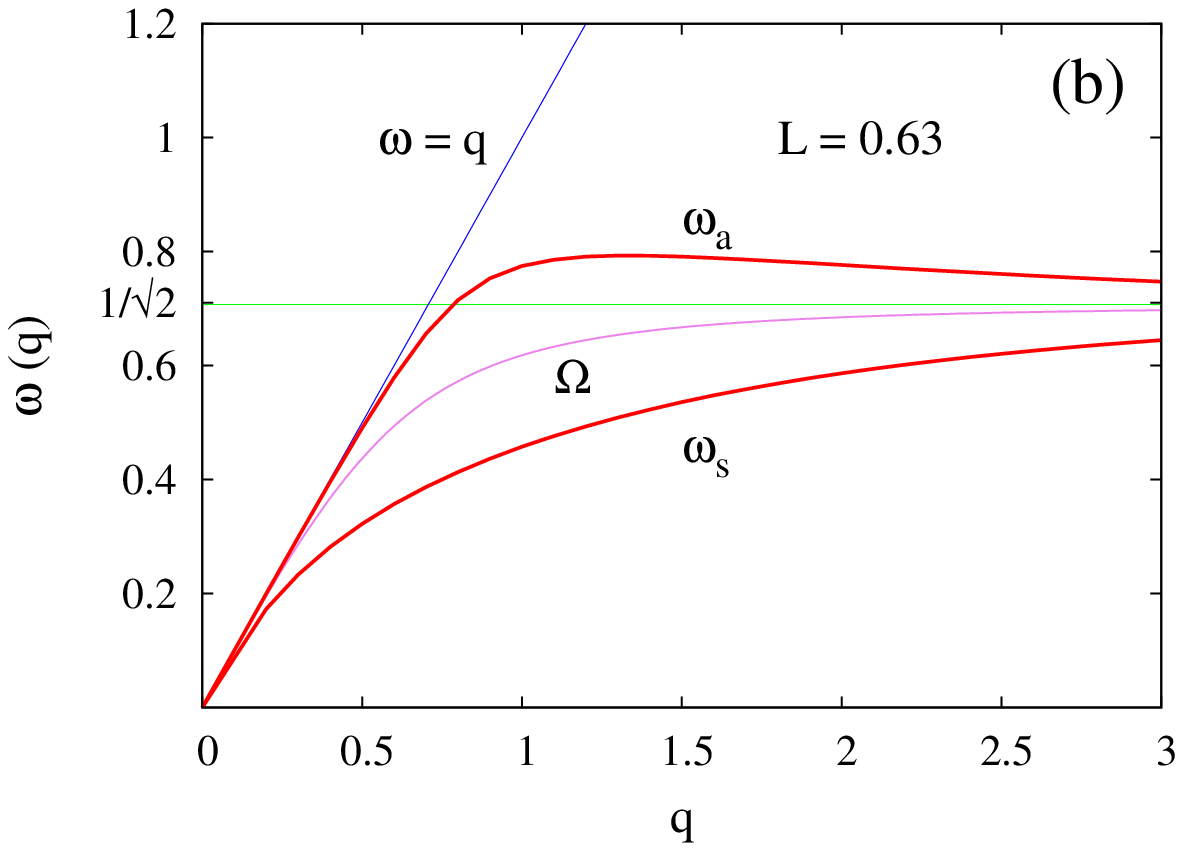}\vskip4mm}\\
\parbox{84.0mm}{\includegraphics[width=84.0mm]{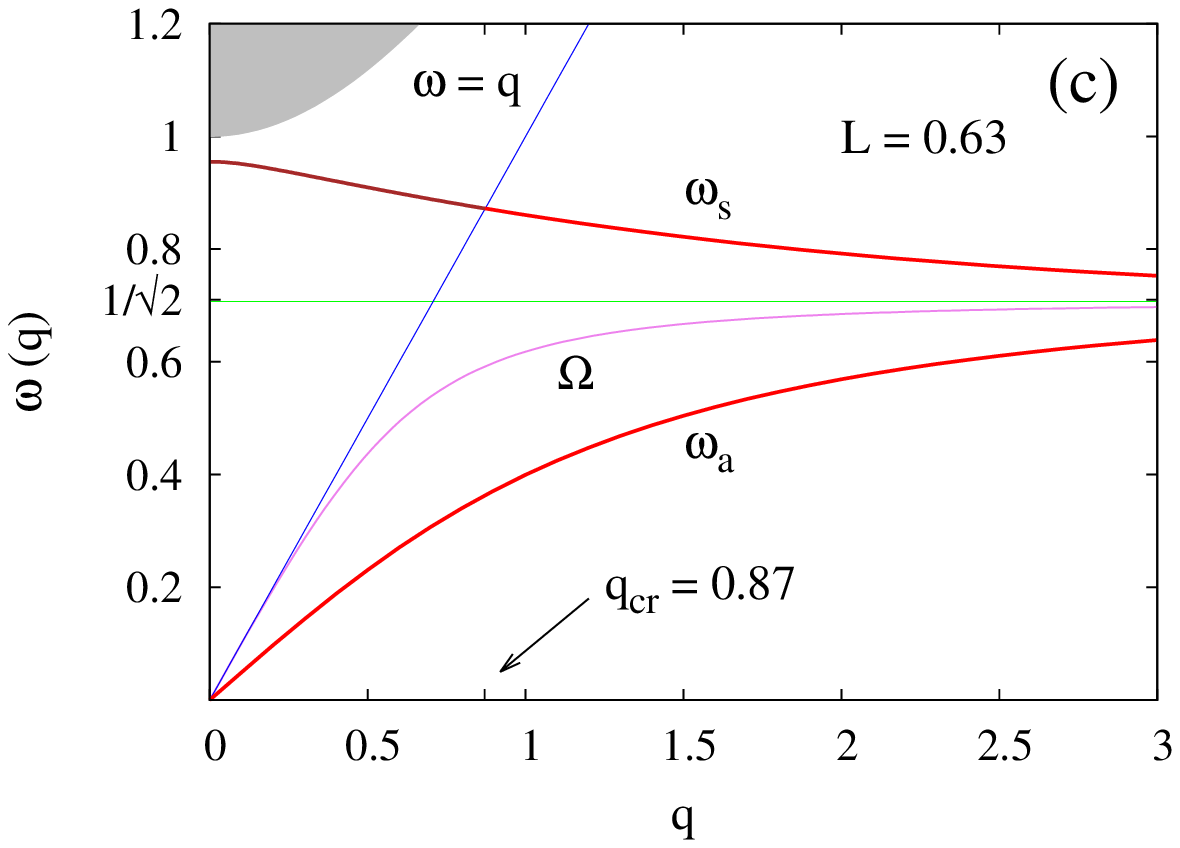}\hfill}&
\parbox{84.0mm}{\hfill\includegraphics[width=84.0mm]{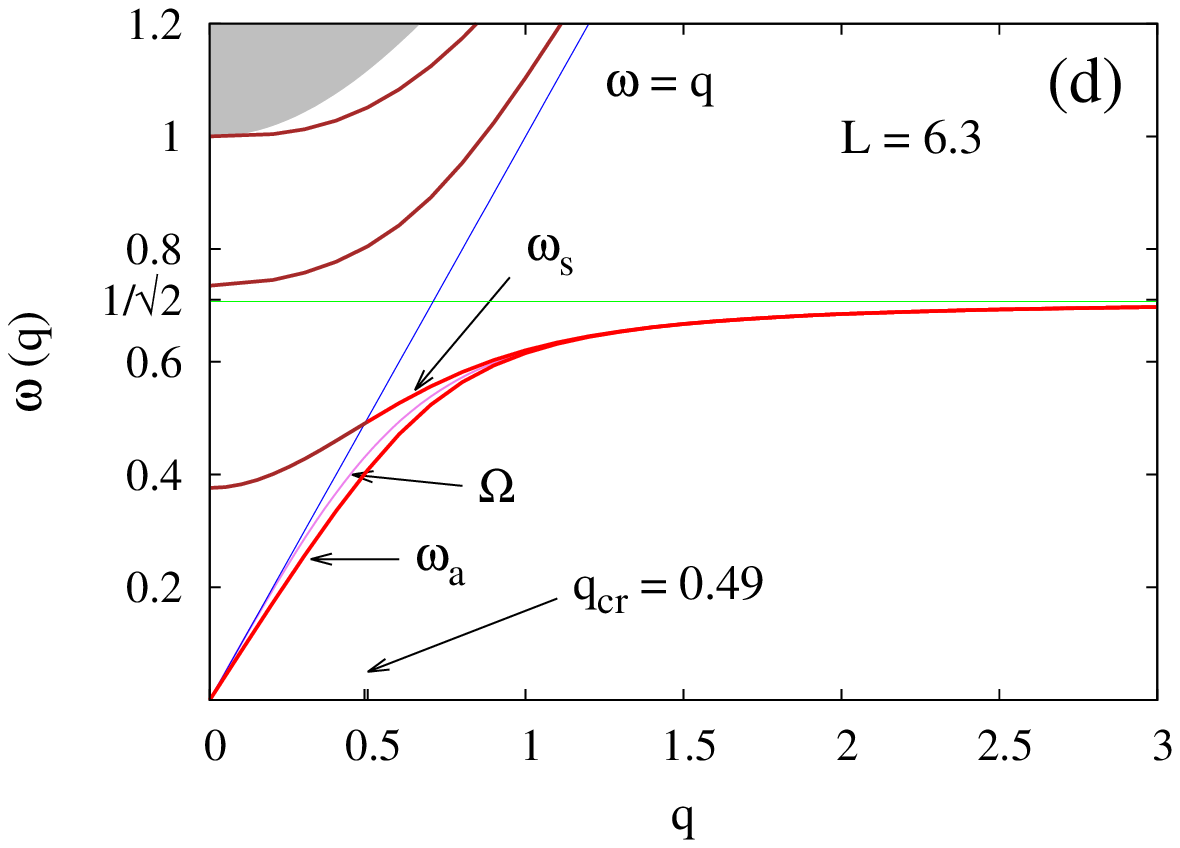}}\\
\end{tabular}}
\caption{\label{Fig1}(Color  online) Dispersion curves for the discrete modes
in the TM polarization for different configurations: (a) single
metal-vacuum interface; (b) metal plate of thickness
$L=2a k_p=0{.}63 $ in vacuum; (c) vacuum gap of the same
width $L=0{.}63$ in the metal bulk and (d)  vacuum gap of
the width $L=6{.}3$ in the metal bulk. Frequencies
$\Omega $,  $\omega$, and the wave number $q$ are presented in the
dimensionless units defined in Eq.\ (\ref{2-10}). The gray
filled regions correspond to the continuous spectrum.}
\end{figure}

This condition  is fulfilled, for instance, at the
metal-vacuum interface, provided the dielectric function of
the metal $\varepsilon_1(\omega)$ is described by the
plasma model (\ref{m}). In this case we put
$\varepsilon_2=\mu_1=\mu_2=1$. Substituting all this into
(\ref{2-5}) and (\ref{2-7}), one obtains the dispersion
equation for the surface plasmon in the TM-polarization
\begin{equation}
\label{2-9} \Omega^4-(1+2q^2)\,\Omega^2+q^2=0\,{,}
\end{equation}
where the dimensionless variables
\begin{equation}
\label{2-10} \Omega=\frac{\omega}{\omega_p}\,{,} \quad
q=\frac{k}{k_p}\,{,}\quad k_p=\frac{\omega_p}{c}
\end{equation}
are introduced. The solution to Eq.\ (\ref{2-9}) is given by
\begin{equation}
\label{2-11}
\Omega(q)=\sqrt{\frac{1}{2}+q^2-\sqrt{\frac{1}{4}+q^4}}\,{.}
\end{equation}
The dispersion curve $\Omega(q)$ is presented in
Fig.~\ref{Fig1}(a). Another solution to Eq.\ (\ref{2-9})
leads to positive $k_1^2$ that corresponds to propagating
waves in medium 1 [dotted curve
in  the upper-left corner in  Fig.~\ref{Fig1}(a)]. In view of the
asymptotic behavior
\begin{equation}
\label{2-12}
\Omega(q)\to q \left ( 1-\frac{q^2}{2} + \cdots \right ),\quad
q\to 0, \quad
\Omega(q)\to \frac{1}{\sqrt 2}\left (
1-\frac{1}{8q^2} + \cdots \right ),\quad q\to \infty \,{,}
\end{equation}
the dispersion curve $\Omega (q)$ tends in these limits  to
two straight lines $\Omega =q$  and $\Omega =1/\sqrt 2$
from below. Here $\Omega =1/\sqrt 2$ is the dimensionless
frequency of the surface plasma oscillations  [see Fig.\
\ref{Fig1}(a)].

With  $\mu_1=\mu_2=1$, obviously, there is no surface modes
in the  TE polarization.

The surface waves can be sustained by the dielectric-vacuum
interface too, as the dielectric permittivity, with account for
the dispersion (\ref{d}), may take on negative values. The
corresponding dispersion law can be found, for example,
in~\cite{Polaritons}.

\paragraph{\label{NM} Normal modes for two parallel plane interfaces}
Now let us consider two interfaces parallel to the $xy$
plane and cutting the axis $z$ at the points  $z=-a$ and
$z=a$. For simplicity, we consider from the beginning the symmetric
configuration that can be symbolically represented as
\[
(\varepsilon_1,\mu_1\mid  \varepsilon_2,\mu_2\mid
\varepsilon_1,\mu_1)\,{.}
\]
For a given polarization,  the normal modes can be
subdivided into symmetric  or asymmetric
solutions. Thus we have for the TM-modes
\begin{equation}
\label{2-13} \Phi(z)
=\begin{cases} A\,e^{-ik_1(z+a)},& z<a;
\cr B\; {\cos{k_2z}\choose  \sin{k_2 z}}, & -a<z<a; \cr \pm
A\, e^{ik_1(z-a)},& z>a\,{,}\cr
\end{cases}
\end{equation}
and the analogous ansatz for the function $\Psi(z)$ in the case of the TE-modes.
The wave vectors $k_1$ and $k_2$ in (\ref{2-13}) are defined, as before,
by Eq.\ (\ref{2-5}).
For $|z|>a$ we again consider the  ``outgoing waves''  with
the imaginary wave vector $k_1^2<0$, while in the region $|z|< a$ the wave
vector $k_2$ may be both imaginary, $k_2^2<0$, or real,
$k_2^2>0$. In  the first case, the waves are just surface ones
localized near the interfaces and  in the second case,
they are the waveguide solutions\footnote{In
\cite{IHL}, such solutions are referred to as the cavity
modes.} describing the standing waves between the
interfaces and the evanescent waves outside this region.
Indeed, both solutions are normal modes because they are
squire-integrable on the whole axis~$z$.

The matching conditions at  points $z=-a$ and $z=a$
result in the respective frequency equations. For the
TM modes these equations read
\begin{eqnarray}
L_a&\equiv& i k_1 \varepsilon_2 +\varepsilon_1 k_2 \tan (ak_2)=0
\quad \text{(asymmetric modes)},
\label{2-14}\\
L_s&\equiv& i k_1 \varepsilon_2 -\varepsilon_1 k_2 \cot (ak_2)=0
\quad \text{(symmetric modes)} \label{2-15}{.}
\end{eqnarray}
When defining the spatial symmetry of the electromagnetic
modes we follow the papers
\cite{Raether-1,Raether-2,Economou}, namely, the symmetric
TM modes are constructed by making use of the symmetric
function $\Phi'(z)$ and, consequently,   antisymmetric
function $\Phi(z)$ in Eqs.\ (\ref{a-12}) and (\ref{a-13}). It
implies that  the longitudinal component of the electric field
$E_x(z)$ is symmetric with respect to the plane $z=0$ and
the transverse components of the fields $E_z(z)$ and
$H_y(z)$ are antisymmetric. The antisymmetric TM modes have
the contrary symmetry properties, i.\ e.\ $E_x(z)$ is
antisymmetric with respect to the $z=0$ plane, and $E_z(z)$
and $H_y(z)$  are symmetric. In the literature
\cite{Vainshtein,Burke} reverse notation is used too.
The eigenfrequencies of the
symmetric modes are denoted by $\omega_s(k)$ and those for the
antisymmetric modes by $\omega_a(k)$. In the literature different
notation for these frequencies is used, for example, sometimes $\omega^+(k)$  is
referred to the higher frequency modes and  $\omega^-(k)$ applies to
the lower frequency collective oscillations
\cite{Raether-1,Raether-2}.

By making use of the trigonometric relations
\[
\tan 2x=\frac{2\tan x}{\tan^2x-1}=\frac{2\cot x}{\cot^2x-1}\,{,}
\]
the frequency equations (\ref{2-14}) and (\ref{2-15}) can
be combined into one equation
\begin{equation}
\label{2-16}
\tan (2a k_2)=-\frac{2i\eta}{1+\eta^2},\quad\eta =\frac{\varepsilon_2k_1}{\varepsilon_1 k_2}\,{.}
\end{equation}
The frequency equations (\ref{2-14}) and (\ref{2-15}) can also
be rewritten in the exponential form
\begin{equation}
\label{2-17}
1+r_{12}e^{2iak_2}=0 \quad (L_a=0 ),
\end{equation}
\begin{equation}
\label{2-18}
1-r_{12}e^{2iak_2}=0 \quad (L_s=0).
\end{equation}
Multiplying together these two equations we arrive at the
exponential form for the total frequency equation (\ref{2-16})
\begin{equation}
\label{2-19}
D(\omega)\equiv 1-r^2_{12}e^{4iak_2}=0 \quad (L_a L_s=0),
\end{equation}
where the notation
\begin{equation}
\label{2-20}
r_{12}\equiv\frac{\varepsilon_2 k_1-\varepsilon_1 k_2}{\varepsilon_2 k_1+
\varepsilon_1 k_2}=\frac{\eta-1}{\eta+1}
\end{equation}
is introduced. It will be shown later  [see Eq.\
(\ref{a-20})] that $r_{12}$ is the reflection amplitude for a single interface.
The frequency equations for the TE polarization are obtained from
the above equations by replacement
 $\varepsilon_\alpha
\to \mu_\alpha, \quad \alpha=1,2 $.

To comprehend the structure of the discrete spectrum in the
presence of two interfaces, we address  typical examples.
The solutions to the frequency equations (\ref{2-14}) and
(\ref{2-15}) are studied in detail in the theory of dielectric
waveguides~\cite{Vainshtein} and optical
waveguides~\cite{Cheo,Marcuse}, in the theory of surface
plasmons
\cite{Polaritons,Raether-1,Raether-2,Sarid,Sernelius-book,Pitarke} and so
on. However, in these applications the roots of the
frequency equations are often sought  in the form $k =
k(\omega)$ (wave vector as function of frequency). We are
interested in  the inverse function $\omega=\omega (k)$.

While studying the frequency equations (\ref{2-14}) and
(\ref{2-15}) their plots  in terms of the variables $k_1$
and $k_2$ instead of the initial variables $k$ and $\omega$
are helpful and transparent.  To this end, the considered
frequency equation is appended with the relation between
$k_1^2$ and $k_2^2$ following from their definition
(\ref{2-5}). At fixed  $\omega$ one obtains two equations
connecting $k_1$ and $k_2$. The intersection points of the
curves defined by the left-hand sides of these equations
give the solutions of the initial frequency equation (see, for
example, \cite{Vainshtein,MSU}).

Let us briefly describe  the structure of the discrete
spectrum of electromagnetic excitations for some
configurations. Not all these examples can have a direct
relation to the experimental studies of the Casimir forces.
Nevertheless, all of them turn out to be instructive in
revealing the general properties of the electromagnetic
spectrum for two plane parallel interfaces.

We specify {\bf the configuration I} in the following way:
\begin{equation}
\label{2-20a}
\varepsilon_1=\mu_1=1, \quad
\varepsilon_2  \text{ and }\mu_2 \text{ are some constants }
\end{equation}
(a material plate in vacuum without dispersion). This
configuration is studied in detail  in radio engineering as
the  simplest example of a waveguide \cite{HdP1,Vainshtein}. When the conditions
\begin{equation}
\label{2-39-1}
\varepsilon_2>0,\quad \mu_2 > 0,\quad \varepsilon_2\mu_2 >1
\end{equation}
or
\begin{equation}
\label{2-39-2}
\varepsilon_2<0,\quad \mu_2 < 0,\quad \varepsilon_2\mu_2 >1
\end{equation}
are satisfied, there exist waveguide solutions in the form
(\ref{2-13}) with $k_1^2< 0$ and $k_2^2>0$ for both the  TE and
TM polarizations.  For a given value of $k^2$, the  number of such
solutions is finite and is increasing with  $\omega$.  The
respective  real eigenfrequencies lie in  the interval $(c_2k,\,c_1
k)$ with $c_1=c$.

As expected, the surface modes $(k_1^2<0,\;
k_2^2<0)$ in the present configuration exist only under the conditions
(\ref{2-39-2}), in other words, if the plate is made of the
left-handed material (both $\varepsilon_2$ and $\mu_2$ are negative
\cite{Veselago,Shadrivov}). In the general case, the number
of  modes of this kind is equal to 4: there are two surface modes with
the TE polarization (symmetric oscillation and antisymmetric
oscillation) and two analogous modes with the TM polarization.

For symmetric surface modes it is enough to fulfill the condition (\ref{2-39-2}). For antisymmetric surface modes,
one has to impose, in addition to (\ref{2-39-2}), the
following conditions:
\begin{equation}
\label{2-39-3}
0<\frac{a \omega}{c}\,(\varepsilon_2\mu_2-1)<
\begin{cases} |\varepsilon_2|^{-1} & \text{TM-polarization;} \cr
|\mu_2|^{-1} & \text{TE-polarization.}\cr
\end{cases}
\end{equation}
The real frequencies of the surface modes lie in the interval  $(0,\,c_2k)$.

As the next example we consider {\bf configuration II}
which is complementary to the previous one, namely,
$\varepsilon_1$  and $\mu_1$ are some positive or negative constants such that
$\varepsilon_1\mu_1>1$ and
$\varepsilon_2=\mu_2=1$
(two semi-spaces of the same material without dispersion
separated by the vacuum gap).

This configuration is not investigated in radio
engineering and waveguide optics,  but it is closely
related to the Casimir calculations. In the case of
left-handed material obeying the conditions
\begin{equation}
\label{2-39-4}
\varepsilon_1<0, \quad \mu_1<0, \quad \varepsilon_1\mu_1>1{,}
\end{equation}
there exist surface symmetric and antisymmetric waves in
both polarizations~\cite{Shadrivov,PL}. Indeed, one of the branches
of the hyperbola defined by the equation
\begin{equation}
\label{2-39-5}
a^2\kappa_2^2-a^2\kappa_1^2=\frac{a^2\omega^2}{c^2}(\varepsilon_1\mu_1-1)>0\,{,}
\quad k_\alpha =i\kappa_\alpha,\quad \alpha=1,2
\end{equation}
at negative  $\varepsilon_1$ and $\mu_1$ always crosses in
the first quadrant the curves defined by the equations
\[
a\kappa_1=-\frac{\varepsilon_1}{\varepsilon_2}(a\kappa_2)\tanh(a\kappa_2)\,{,}
\quad
a\kappa_1=-\frac{\varepsilon_1}{\varepsilon_2}(a\kappa_2)\coth(a\kappa_2)
\]
and two more obtained by replacing  $\varepsilon_\alpha$ by
$\mu_\alpha, \quad \alpha=1,2$. Therefore, the discrete spectrum for
the present configuration exists only in the case of
left-handed materials and consists of four surface modes. The
frequencies of these modes belong to the interval  $0<
\omega< c_1k$.

Keeping in mind  the possibility of total reflection at the
interface of the media 1 and 2 one can try to find the
solutions for this configuration with  $k_1^2>0$ and
$k_2^2<0$. However, it is easy to see that there are no such
solutions with real frequencies as the equations are
complex. Even if such solutions existed, they are not
eigenfunctions of the discrete spectrum because they are
not squared integrable on the infinite axis  $z$.

For the configuration under consideration there are no
waveguide solutions to the Maxwell equations. Indeed, for
the existence of such solutions, it is required that $k_1^2<0$,
and $k_2^2>0$. From these conditions we arrive at
the conflicting inequalities $0<\omega<c_1k$ and $\omega> ck$,
with $c_1=c/\sqrt{\varepsilon_1\mu_1}<c$.

Now let us consider the examples of the discrete
electromagnetic spectrum  explicitly allowing for the
temporal dispersion in the media. As {\bf configuration III} we take
a material (metal) slab in vacuum, the dielectric properties of the
slab being described by the simple plasma model
(\ref{m}) [see Fig.\ \ref{Fig1}(b)]:
\begin{equation}
\label{2-39-4a}
\varepsilon_1=\mu_1=1, \quad \varepsilon_2(\omega)=
1-\frac{\omega_p^2}{\omega^2}, \quad \mu_2=1\,{.}
\end{equation}

This configuration supports two surface waves in the TM-polarization. These modes are well studied theoretically
\cite{Economou,Shadrivov} and experimentally, for instance,
in  measurements of the energy losses  by the electrons passing
through the metal films  \cite{Raether-1,Raether-2}.
There is vast literature on this subject
\cite{Polaritons,Sarid,Pitarke}. Our citing is minimal.

The physical picture of these collective  excitations is
the following. At both boundaries of the metal plate $(z=\pm
a)$ with $a \to \infty$ there  exist plasma oscillations of
the electron density  with the same  frequency $\omega_p/\sqrt{2}$
and  dispersion law (\ref{2-11}). If the thickness of the
plate or film is finite these oscillations interact reciprocally and the energy
level (\ref{2-11}) splits, as usual, in two levels (asymmetric
and symmetric ones) with frequencies $\omega_a(k)$ and
$\omega_s(k)$ defined by  Eqs.\  (\ref{2-14}) and
(\ref{2-15}), respectively [see Fig.\ \ref{Fig1}(b)].

It is important that both curves $\omega_s(k)$ and
$\omega_a(k)$ are placed  to the right of the line  $\omega=ck$ and
below the horizontal line $\omega = \omega_p$. Thus the
eigenfrequencies of the surface modes lie in the interval
$0\leq \omega_s(k),\,\omega_a(k)< ck$ for $0<k<\infty$.

The curve $\omega_{a}(k)$ at all  $k$ lies above the curve
$\omega_{s}(k)$. Near the origin  both the  curves
touch the line $\omega=ck$ from the right. With  $k \to
\infty$ the curves  $\omega_s(k)$ and $\omega_a(k)$ tend to
the horizontal line  $\omega = \omega_p/\sqrt{2}$, respectively,
from below and from above.  For $2a \to \infty$ the
frequencies  $\omega_{s}(k)$  and $\omega_{a}(k)$ tend to
the universal curve  (\ref{2-11}) from below and from
above.

The characteristic value of the longitudinal wave vector
$k$  in the problem under consideration is
$k_p=\omega_p/c$. It is within the vicinity of $k \sim k_p$
where the ``bend'' of the curves $\omega_s(k)$ and
$\omega_a(k)$ is observed.

It is evident that there are no waveguide solutions
$(k_1^2<0,\; k_2^2>0)$ in the present  case. Indeed, from
the definition of the wave vectors
 $k_1$ and $k_2$ (see Eqs.\ (\ref{2-5})
 and  (\ref{m})) we obtain
\[
            -k_1^2+k_2^2=-\frac{\omega_{p}^2}{c^2}\,{.}
\]
For the waveguide solutions the left-hand side of this
equation should be positive. Hence, this equality cannot be
satisfied.

Now we consider {\bf configuration IV}, complementary to
the previous one, namely, a vacuum slot, $2a$ in width,
in the bulk metal described by the plasma model (see Fig.\ \ref{Fig1}c,d)
\begin{equation}
\label{2-39-6}
\varepsilon_1(\omega)= 1-\frac{\omega^2_p}{\omega^2},\quad \mu_1=1,\quad
\varepsilon_2=\mu_2=1\,{.}
\end{equation}
First, we address  the surface waves in the present set-up,
$k_1^2<0,\;k_2^2<0$, which exist in the TM polarization
\cite{Economou}. Again, we have two such collective modes,
symmetric and antisymmetric ones, their frequencies being
placed in the interval $(0,ck)$. The lay-out of the
dispersion curves is interchanged in comparison with the
previous case, $\omega_s(q)>\omega _a(q)$. The curve
$\omega _a(q)$ lies below the horizontal line $\omega =
1/\sqrt{2}$ and to the right from the line $\omega=q$. At
the point $q=q_{\text{cr}}=1/\sqrt{1+L/2}$ the dispersion
curve $\omega _s(k)$  crosses the light line $\omega = q$
and becomes the lowest waveguide mode (see
Figs.~\ref{Fig1}(c) and \ref{Fig1}(d) and
\cite{Economou}).\footnote{Probably, this implies that  the
symmetric surface TM mode in this configuration may directly be
excited by the light beam contrary to  the plate
configuration  [Fig.~\ref{Fig1}(b)]
\cite{Raether-1,Raether-2,Sarid,Polaritons}.} Thus, the
symmetric surface mode $\omega_s(q)$ becomes in this
configuration a ``hybrid'' one. It is this fact that leads to
the known result, namely, the contribution of the surface modes
to the Casimir energy for a given configuration proves to be
considerable only at small distances. Indeed, because of
the hybrid nature of the symmetric mode, the dispersion
curves $\omega_s(q)$  and $\omega_a(q)$ move apart in the
region $q<q_{\text{cr}}$ [Fig.\ \ref{Fig1}(c) and \ref{Fig1}(d), this
divergence being maximum at small $L$ [compare Figs.\
\ref{Fig1}(c) and  \ref{Fig1}(d)]. Obviously, this implies
in turn that the mutual interaction of these collective excitations
is maximum at small $L$.

With increasing the gap width $L$, in this configuration
there appear  waveguide  solutions $(k_1^2<0, \; k_2^2>0)$
satisfying the conditions
\begin{equation}
\label{2-40}
c^2 k^2<\omega^2_{wg}(k)< \omega^2_p+c^2 k^2\,{.}
\end{equation}
Thus, they lie to the left from the line  $\omega=ck$ and below the curve
$\omega= \sqrt{\omega^2_p+c^2 k^2}$.
Approximately,  the frequency of these modes and their total numbers, for each polarization,
are given by~\cite{Chang}
\begin{equation}
\label{2-41}
\omega^2(k)\approx k^2 c^2+n^2\pi^2c^2/4a^2,
\quad n=0, 1,2, \ldots , n< 2a\omega_p /\pi c\,{.}
\end{equation}
These modes exist only at large gaps  (compare
Fig.~\ref{Fig1}c and \ref{Fig1}d). Indeed, for gold
$\omega_p=1{.}38\cdot10^{16}$~s$^{-1}$. Therefore, the
dimensionless width of the gap $L=0{.}63$ implies
$2a=14$~nm and $L=6.3$ corresponds to $2a=140$~nm. In the
first case, there are no pure waveguide modes
[Fig.~\ref{Fig1}(c)] and, at the same time, the surface modes
interact strongly because their dispersion curves are
distinguished considerably. In the second case
[Fig.~\ref{Fig1}(d)], only two pure waveguide modes
appear while the surface modes at $q>1$ practically
coincide with each other and with the single interface
dispersion  curve $\Omega (q)$. The latter means that in
this region the surface modes do not couple. For copper, we
have close figures: $\omega_p=1{.}46\cdot
10^{16}$~s$^{-1}$, $2a=12{.}6$~nm ($L=0{.}63$), and
$2a=126$~nm ($L=6{.}3$).

The fact that waveguide modes exist only for large gaps
suggests that they contribute appreciably to the dispersion
force at large distances~\cite{Chang}, where contributions
of the surface modes practically vanish.

The examples studied above reveal  general properties of
the discrete spectrum in the problem at hand and permit us to
establish the lower boundary of the continuous spectrum.
Let us summarize these properties:
\begin{enumerate}
\item The discrete spectrum may comprise the solutions of two types:
the surface modes and the wave guide modes;
\item At fixed $k$ the frequencies of the surface modes
(if they exist) lie in the interval
$(0, \omega_{-}(k))$, while the frequencies of the
waveguide solutions (if they exist) belong to the interval
$(\omega_-(k), \omega_+(k))$, where the boundary values
$\omega_-(k)$ and $\omega_+(k)$ ($\omega_-(k)<\omega_+(k)$)
are defined by the explicit form of the dielectric function
$\varepsilon (\omega)$;
\item The lower boundary of the continuous spectrum is defined by the condition
$k_1^2(\omega)>0$.
\end{enumerate}

Here we list  the values of $\omega_+(k)$ and  $\omega_-(k)$
for the configurations I -- IV.
\[
\text{I:}\quad \omega_-(k)=c_2 k\equiv
\frac{ck}{\sqrt{\varepsilon_2\mu_2}}\,{,} \quad
\omega_+(k)=ck,\quad c_2<c\,{.}
\]
In this case, the discrete spectrum may include the surface modes and the
waveguide modes with the frequencies from the above-specified
intervals; the continuous spectrum $\omega$ is defined by
$\omega>\omega_+(k)$.
\[
\text{II:}\quad \omega_-(k)=c_1 k\equiv
\frac{ck}{\sqrt{\varepsilon_1\mu_1}}\,{,} \quad
\omega_+(k)=ck,\quad c_1<c\,{.}
\]
The discrete spectrum may consist  of the surface modes only
with the frequencies from the above-specified interval; the
continuous spectrum $\omega$ exists for   $\omega>\omega_-(k)=c_1
k$.
\[
\text{III:}\quad \omega_-(k)=c k\,{,} \quad
\omega_+(k)=\sqrt{c^2k^2+\omega_p^2}\,{.}
\]
The discrete spectrum is formed by the surface modes with
frequencies from the interval specified  above;   the
continuous spectrum is defined by $\omega>\omega_-(k)=c k$.
\[
\text{IV:}\quad \omega_-(k)=c k\,{,} \quad
\omega_+(k)=\sqrt{c^2k^2+\omega_p^2}\,{.}
\]
The discrete spectrum comprises the surface modes and the
waveguide modes with frequencies in the corresponding
intervals;  the continuous spectrum exists for
$\omega>\omega_+(k)=\sqrt{c^2k^2+\omega_p^2}$.

In what follows it is significant that the points
$\omega_-(k)$ and  $\omega_+(k)$ are the branch points
for the root functions $k_1(\omega)$ and $k_2(\omega)$ (see
Sec.\ \ref{Transition}).

The configurations considered above are unbounded, i.e. open. As a
result, in addition to the real roots the frequency equations have
also the discrete complex roots $\omega (k)=\omega'(k)+i\omega''(k)$
which correspond to the quasi-normal modes~\cite{NVV-bc}. These modes are
not squared integrable and therefore they are not normalizable.  That
is why the quasi-normal modes are not included in the completeness
relation. However, their existence should be taken into account when
choosing the contours for integration in the complex $\omega$ plane
(see Sec.\ \ref{Transition}).

It is worth noting here that the quasi-normal modes are
explored in fact in the experiments dealing with the radiating plasmons
\cite{Raether-2}. Usually, in this field the complex wavenumbers are
considered $k(\omega)=k'(\omega)+ik''(\omega)$ instead of the complex
eigenfrequencies.

\subsubsection{\label{Continuous} Treatment of the continuous part of the spectrum}
To  study the contribution of the continuous spectrum  to
the vacuum energy, the scattering matrix is required, i.e.,
the  $S$ matrix for the wave equations (\ref{a-14}) and
(\ref{a-15}) with the   matching conditions (\ref{a-16})
and  (\ref{a-17}), respectively. Here we  deal  with
the one-dimensional scattering on the  infinite $z$ axis, $-\infty<
z< \infty$. This problem has some peculiarities
\cite{Fluegge,Messiah,Barton,Newton}. Specifically, one has
to consider the  incidence of the initial waves from the left
and from the right separately. Thus,  we must add such
initial waves to the outgoing and standing waves in Eqs.\
(\ref{2-4}) and (\ref{2-13}) considered in  the pertinent
eigenvalue problems.
The respective
reflection amplitudes, $r_l,\,r_r$, and transition amplitudes,
$t_l,\,t_r$, make up  a ($2\times 2$) matrix that plays the role
of the scattering matrix in the problem under consideration
\begin{equation}
\label{a-19} S(\omega)=\left ( \begin{array}{cc}
t_l & r_l \\
r_r & t_r
\end{array}
\right ){.}
\end{equation}
\paragraph{Single plane interface}
    In the case of the one-plane interface $(\varepsilon_1,\mu_1\mid \varepsilon_2,\mu_2)$
we have for the TM polarization,
\begin{eqnarray}
r_l\equiv r_{12}=\frac{\varepsilon_2 k_1-\varepsilon_1 k_2}{\varepsilon_1 k_2+
\varepsilon_2 k_1},\quad  & r_r\equiv r_{21}=-r_{12}=-r_l\,{,} \label{a-20}\\
t_l\equiv t_{12}=\frac{2\varepsilon_1k_1}{\varepsilon_1 k_2+
\varepsilon_2 k_1},\quad  & t_r\equiv t_{21}=\frac{\displaystyle
2\varepsilon_2 k_2}{\displaystyle \varepsilon_1 k_2+
\varepsilon_2 k_1}, \quad t_{12}\neq t_{21}\,{.}\label{a-21}
\end{eqnarray}
The analogous amplitudes for the TE-polarization are obtained
from Eqs.\ (\ref{a-20}) and (\ref{a-21}) by the substitution
$\varepsilon_\alpha\to \mu_\alpha,\; \alpha=1,2$.

The amplitudes $t$ and $r$ for the TM and TE
polarizations have the poles at the points where the
variable $\omega$ coincides with the eigenfrequencies
defined by Eqs.\ (\ref{2-7}) and  (\ref{2-8}). This
property is preserved for several interfaces too (see below).
The single-plane interface is a  nonsymmetric configuration;
therefore, the $S$ matrix (\ref{a-19})  in this  case does not possess
appealing physical properties.

\paragraph{Electromagnetic scattering by two parallel plane interfaces}

Now we address the scattering matrix for two  plane
interfaces which are parallel to the $xy$ plane and cut the  $z$
axis at the points $z=-a$ and $z=a$, in other words, we
envisage the configuration $(\varepsilon_1,\mu_1\mid
\varepsilon_2,\mu_2\mid \varepsilon_3,\mu_3)$.  The
explicit form of $t$ and  $r$ is again derived from the
matching conditions (\ref{a-16}) and (\ref{a-17})
\cite{Stratton,Born-Wolf} or by making use of the multiple
reflection technique \cite{Born-Wolf,Brevik}. By taking
careful account of the phase factors at the points $z=-a$
and $z=a$ one obtains
\begin{equation}
\label{a-22}
r_l\equiv r_{123}=\frac{r_{12}+r_{23}e^{4ik_2a}}{1+r_{12}r_{23}e^{4ik_2a}}\,{,}
\quad t_l\equiv t_{123}=\frac{t_{12}t_{23}e^{2ik_2a}}{1+r_{12}r_{23}e^{4ik_2a}}\,{,}
\end{equation}
\begin{equation}
\label{a-23}
r_r\equiv r_{321}=\frac{r_{32}+r_{21}e^{4ik_2a}}{1+r_{32}r_{21}e^{4ik_2a}}\,{,}
\quad t_r\equiv t_{321}=\frac{t_{32}t_{21}e^{2ik_2a}}{1+r_{32}r_{21}e^{4ik_2a}}\,{,}
\end{equation}
where $r_{ij}$ and $t_{ik}$ are defined in Eqs.\ (\ref{a-20}) and (\ref{a-21}).

Furthermore, we consider the special case of  (\ref{a-22}) and
(\ref{a-23}), the symmetric one,
$(\varepsilon_1,\mu_1 \mid \varepsilon_2,\mu_2
\mid\varepsilon_1,\mu_1 )$.  In the symmetric setup one  has
\begin{equation}
\label{a-24}
r=r_l=r_r\equiv r_{121}=\frac{r_{12}-r_{12}e^{4ik_2a}}{1-r^2_{12}e^{4ik_2a}}\,{,}
\quad t=t_l= t_r\equiv t_{121}=\frac{t_{12}t_{21}e^{2ik_2a}}{1-r^2_{12}e^{4ik_2a}}\,{.}
\end{equation}

The amplitudes $r$ and $t$ in Eq.\ (\ref{a-24}) have the
poles at the points of the discrete spectrum [see Eq.\
(\ref{2-19})], and they obey the relations
\begin{eqnarray}
|t|^2+|r|^2&=&1,\label{a-25a} \\ r\bar t+\bar r t&=&0\,\label{a-25b}{,}
\end{eqnarray}
where the bar means complex conjugation. Now the scattering
matrix (\ref{a-19}) is given by
\begin{equation}
\label{a-26} S(\omega)=\left ( \begin{array}{cc}
t & r \\
r & t
\end{array}
\right ){.}
\end{equation}
Due to Eqs.\ (\ref{a-25a}) and (\ref{a-25b}) the $S$ matrix
(\ref{a-26}) is a symmetric unitary matrix
\begin{equation}
\label{a-27}
SS^{\dag}=S^{\dag}S =1\,{.}
\end{equation}

In the case of potential scattering on an infinite line
with symmetric potential the amplitudes  $t$ and $r$
are parametrized as follows~\cite{Barton}:
\begin{equation}
\label{2-28}
t=\cos\theta \,e^{i\delta},\quad r=i\sin \theta \,e^{i\delta}\,{,}
\end{equation}
where $\theta $ and $\delta $ are the real functions of $\omega $. Further, we have
\begin{subequations}
\label{all-2-29}
\begin{eqnarray}
\det S(\omega)=t^2(\omega)-r^2(\omega)&=&e^{2i\delta(\omega)}\label{2-29a} \\
&=&\frac{t(\omega)}{\bar t(\omega)}{.}\label{2-29b}
\end{eqnarray}
\end{subequations}
For the real  $\omega$ the function $\det S(\omega)$ is defined by  formula (\ref{2-29a}),
while (\ref{2-29b}) gives its continuation to the whole complex plane $\omega$.

In order to consider in this approach the initial
asymmetric configuration $(1|2|3)$, one has to proceed from
the symmetric multilayered configuration $(1|2|3|2|1)$,
calculate the respective total reflection and transition
amplitudes, and find $\det S(\omega)$. Transition to the
initial asymmetric configuration should be done only at the
final stage of the vacuum energy  calculation (see Sec.\
\ref{Transition}) by moving the interface $3|2$  to infinity.

\section{\label{Casimir}Casimir energy}
\label{Casimir} We define the Casimir energy as the
renormalized zero-point energy of the electromagnetic
oscillations for the configuration at hand
\begin{equation}
\label{3-1} E = \frac{\hbar }{2}\sum_{\{q\}}
\omega_q\,{.}
\end{equation}
Here the sum  sign implies the summation over the discrete part
of the spectrum and the integration over its continuous branch.

The summing up of the discrete eigenfrequencies will be
performed by making use of the frequency equations. The
integration over the continuous part of the spectrum cannot
be carried out by applying the ``naive'' formula
\begin{equation}
\label{3-2} \int \omega \, d\omega\,{,}
\end{equation}
because it does not ``feel'' the physical nature of the
problem under consideration, namely, this integral does not
depend on the specific form of the differential operator
and the corresponding boundary conditions
~\cite{NVV-bc}. Therefore, the summation
over the continuous branch of the spectrum should be
accomplished, as usual, by making use of the spectral
density shift. The rigorous mathematical theory of
the scattering problem gives the following expression for this
function \cite{Krein}:
\begin{subequations}
\label{all-3-3}
\begin{eqnarray}
\Delta \rho(\omega)\equiv\rho (\omega)- \rho
_0(\omega)&=& \frac{1}{\pi}\frac{d}{d\omega}\delta(\omega)\label{3-3a} \\
&=& \frac{1}{2\pi i}\frac{d}{d\omega} \ln\det
S(\omega)=\frac{1}{2\pi i}\frac{d}{d\omega}\ln\frac{t(\omega)}{\bar t(\omega)}\,{.}
\label{3-3b}
\end{eqnarray}
\end{subequations}
where $S(\omega)$ is the $S$ matrix in the spectral problem
at hand. In Eqs.\ (\ref{all-3-3})  $\rho(\omega )$ is the
density of states for a given potential (or for  given
matching  conditions in the case of the compound media) and
$\rho_0(\omega)$ is the spectral density in the respective
free spectral problem (for the vanishing potential or for
the homogeneous unbounded space). In the case of
one-dimensional scattering the calculation of the spectral
density was considered in \cite{Barton,Barton-1d}. It is
worth noting here that already at the level of the  spectral density
(\ref{all-3-3}) the removing of infinity is carried out,
namely, the contribution to the vacuum energy
generated by the unbounded Minkowski space is subtracted here.

Taking into account all this we can work out Eq.\ (\ref{3-1}) in detail
as follows:
\begin{equation}
\label{3-4} E(2a)=\frac{\hbar
}{2}\sum_{\sigma}\int_{-\infty}^{\infty} \frac{d^2
\mathbf{k}}{(2\pi)^2} \left[ \sum_{n} \omega
_{n}(\sigma,k,a) + \int _{\omega_0}^\infty \omega\,
\Delta \rho (\sigma,\omega,k,a)\,d \omega \right ]-(a\to
\infty)\,{.}
\end{equation}
Here  $\mathbf{k}$ is the wave vector along unbounded dimensions $\mathbf{k}=(k_x,k_y)$;
$\omega _{n}(\sigma,k,a)$ are the  frequencies appertaining to the discrete spectrum;
the function of the spectral density
shift $\Delta \rho (\omega)$ is given by Eqs.\ (\ref{all-3-3}); and $\omega_0$
is the lower bound of the continuous spectrum. Summation over
$\sigma=\text{TE,\,TM}$ in (\ref{3-4}) takes into account
two polarizations of the electromagnetic field.

The further treatment of Eq.\ (\ref{3-4}) is aimed at deriving a
unique integral representation for the contributions of
both the discrete part and the continuous part of the
spectrum, the integration being carried out over the imaginary
frequencies $\omega = i \zeta$.

\begin{figure}[p!]
\noindent \centerline{
\includegraphics[width=100mm]{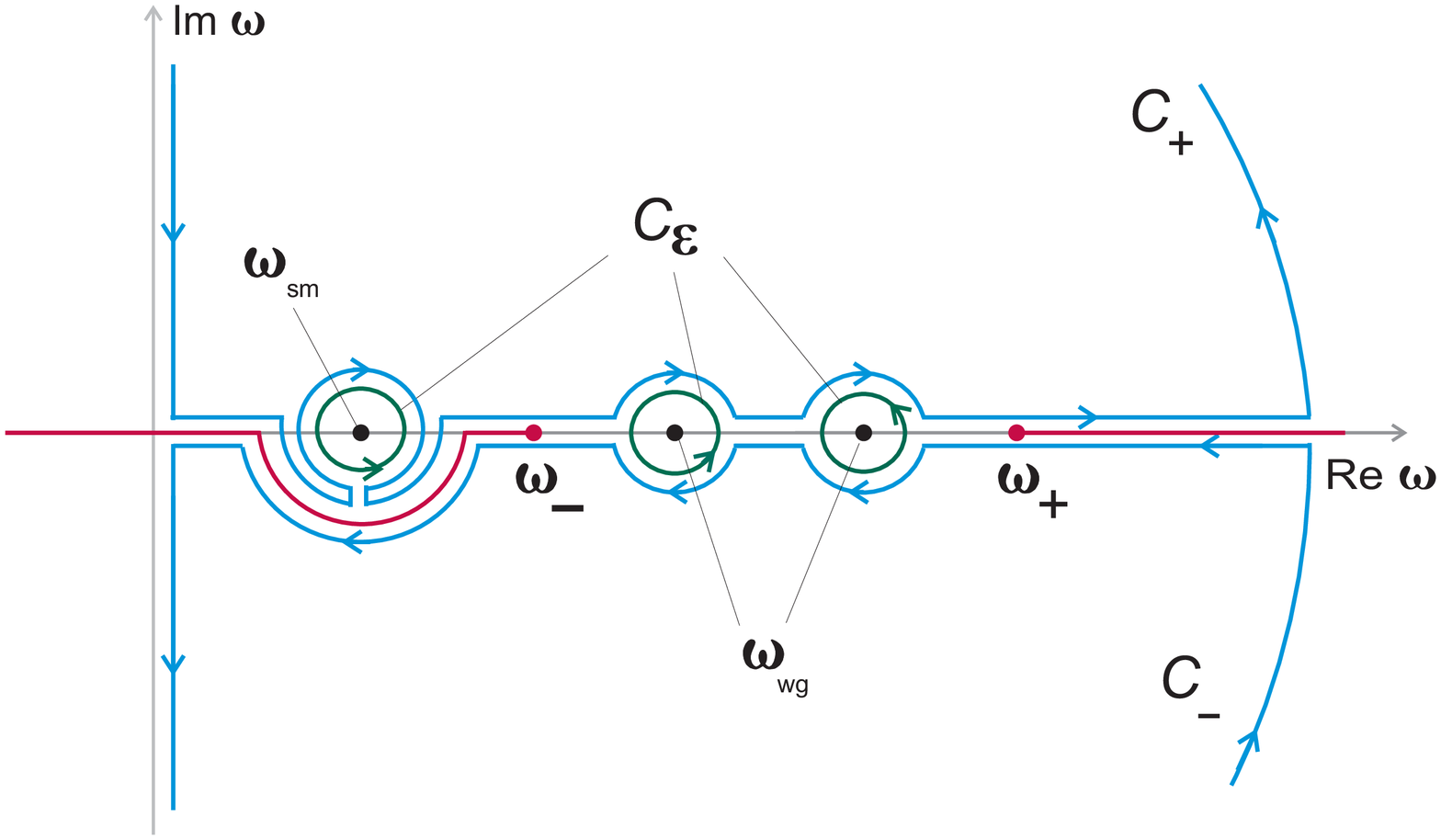}}
\caption{\label{contour}(Color  online) The contours $C_+$ and $C_-$ in the
complex $\omega$ plane which are used when going on to the
integration over the imaginary  frequencies. For
simplicity, the spectrum possessing one surface mode
$\omega_{sm}$, two waveguide modes $\omega_{wg}$, and
the continuous part $\omega > \omega_+$ is considered. The cuts
starting at the points $\omega_-$ and $\omega_+$ are shown
by heavy lines.}
\end{figure}

To be definite, we shall consider the  ``richest''
electromagnetic spectrum possessing the surface modes,
waveguide modes, and the continuous part (see, for example,
configurations I and IV in Sec.\ \ref{Discrete}). As was
shown there, the frequencies of the surface modes lie in
the interval $0<\omega_{sm}<\omega_-(k)$, the frequencies
of the waveguide modes belong to the band
$\omega_-(k)<\omega_{sm}<\omega_+(k)$, and for the frequencies
$\omega $ of the continuous part of the spectrum we
have $\omega_+(k)<\omega$ (see Fig.~\ref{contour}).
\noindent

\subsection{\label{Transition} Transition to imaginary frequencies}
In what follows, we have to go out in the complex frequency plane $\omega$.
For this the one-valued branches of the integrand in Eq.\ (\ref{3-4}) should be
chosen. The square root dependence
\begin{equation}
\label{3-5}
k_\alpha(\omega)=\sqrt{\varepsilon_\alpha(\omega)\frac{\omega^2}{c^2}-k^2},\quad \alpha=1,2
\end{equation}
results in branch points. For the usually accepted functions
$\varepsilon(\omega)$ there appear four such points on the real
axes: $\pm\omega_-(k)$ and $\pm\omega_+(k)$
($0\leq\omega_-(k)<\omega_+(k)$) (see Sec.\
\ref{Discrete}). These are the same points that separate the
different parts of the spectrum.

We draw two cuts on the real axes $\omega$: the first cut
connects the points $-\omega_-$ and $\omega_-$ and the
second one starts at the point $\omega_+$ and goes to
infinity (see Fig.~\ref{contour}).
The first cut  passes round the
eigenfrequency of the surface mode $\omega_{sm}$ from
below. It is not crucial  and this cut can go round
$\omega_{sm}$ from above as well. Further, we assume that $k_1(\omega)$
and $k_2(\omega)$ in Eq.\ (\ref{3-5}) acquire real positive
values at the upper edges of the cuts.

The eigenfrequencies  $\omega_n(\sigma, k, a)$ are unknown
in an explicit form, we have only the respective frequency
equations (\ref{2-14}) and (\ref{2-15}) or in another form
(\ref{2-17}) -- (\ref{2-19}). This dictates, without
choice, the making use of  the argument principle theorem from
the complex analysis when summing over $n$ in (\ref{3-4}).
When choosing the contour for integration in the complex $\omega$
plane one should bear in mind that the left-hand sides of
the frequency equations $D(\sigma,\omega)$ considered as
the functions of the complex variable $\omega$ have a ``good'' behavior
in the upper half plane $\omega$ only:
\begin{equation}
\label{3-6}
D(\sigma,\omega)=1-r_{\sigma}^2e^{4iak_2}, \quad ck_2=\sqrt{\omega^2-\omega_+^2}\,{.}
\end{equation}
This forbids from the very beginning  the deformation of
the contour in such a way that it includes the whole
imaginary axes $\im \omega$. An admissible contour is
obviously the following  one: the integration should be performed
along the circles $C_\varepsilon$  of the radius  $\varepsilon$ with $\varepsilon\to
0$, with each circle surrounding  a single root of the frequency equation
(see Fig.~\ref{contour}).

When treating the contribution of the continuous part of
the spectrum to the vacuum energy one should keep in mind
the following. Upon substituting (\ref{3-3b})  in $\Delta\rho(\omega)$
the transition amplitude from (\ref{a-24}),
\begin{equation}
\label{3-7}
t(\omega)=\frac{t_{12}t_{21}e^{2ik_2a}}{D(\omega)},
\end{equation}
in vacuum  energy (\ref{3-4}) there
appear  terms which are proportional to the distance
between the slabs $2a$. These contributions are generated by
the factor $e^{2ik_2a}$ and they belong to the internal
energy of medium  2 in the configuration under
consideration. These terms are removed from the Casimir
energy by the subtraction indicated in Eq.\ (\ref{3-4}).
Hence, in calculation of the vacuum electromagnetic energy
the denominator $D(\omega)$  in  Eq.\ (\ref{3-7}) ``works'' alone.

Now the vacuum energy (\ref{3-4}) acquires the form
\begin{equation}
\label{3-8} E(2a)=\frac{\hbar
}{2}\sum_{\sigma}\int_0^{\infty} \frac{k
dk}{2\pi}\,\frac{1}{2\pi i} \left [\int_{C_{\varepsilon}} d\omega\,
\omega \frac{d}{d\omega}\ln D(\sigma,
\omega)+\int_{\omega_+}^\infty d\omega\,
\omega \frac{d}{d\omega}\ln \frac{\bar D(\sigma,\omega)}{D(\sigma,\omega)}
\right ]{.}
\end{equation}

Furthermore, we represent the contribution of the continuous
spectrum to the vacuum energy [the second term between the square
brackets in Eq.\ (\ref{3-8})] in the form of the contour
integrals as well. According to the  Cauchy integral
theorem, we can write
\begin{subequations}
\label{all-3-9}
\begin{eqnarray}
\int_{C_+}d\omega\,\omega\frac{d}{d\omega}\ln D(\sigma,\omega) &=&0, \label{3-9a} \\
\int_{C_-}d\omega \,\omega\frac{d}{d\omega} \ln \bar D(\sigma,\omega)&=&0, \label{3-9b}
\end{eqnarray}
\end{subequations}
where the contours $C_+$ and $C_-$ are shown in
Fig.~\ref{contour}. Here we have taken into account that
the complex roots of the equation $D(\sigma, \omega)=0$ lie
only in the lower half-plane $\omega$ and, respectively, the
complex roots of the equation $\bar D(\sigma,\omega)=0$ are
placed only in the upper half-plane $\omega$.\footnote{A
simple example of the complex frequencies in solutions to
the Maxwell equations for an  open system (a perfectly conducting
sphere in the vacuum) was discussed in Ref.\ \cite{NVV-bc}.}
Such a lay-out of the complex eigenfrequencies is in a direct
relation with the choice of the time dependence in the
solutions to the Maxwell equations in the form $e^{-\omega
t}$ (see Sec.~\ref{Maxwell}). In view of   relations (\ref{all-3-9}),
we can express the contribution of the continuous part of
the spectrum in (\ref{3-8}) as the integrals along the
contours $C_+$ and $C_-$ which have no  section
$(\omega_+,\,\infty)$. One can easily see from
Fig.~\ref{contour} that along the contours  $C_\varepsilon$
the contributions of the discrete spectrum and continuous
spectrum are mutually canceled.\footnote{In earlier papers
dealing with other configurations the analogous
cancellation was brought out by making use of the complex
$k_2$ plane  (see, for example, \cite{Bordag-95,Kirsten}).
However, the Lifshitz formula is presented in terms of the
variable $i\zeta=\omega$.} Besides, at  the length $(0,
\,\omega_+)$ the contributions of the continuous part of
the spectrum generated by $D(\sigma,\omega)$ and $\bar
D(\sigma,\omega)$ are mutually canceled too.

As a result, we
arrive at a very compact and clear formula for the Casimir
energy
\begin{equation}
\label{3-10} E(2a)=\frac{\hbar
}{2\pi}\sum_{\sigma}\int_0^{\infty} \frac{k
dk}{2\pi}\int_0^\infty d\zeta\,
\ln D(\sigma, i\zeta,k){,}\quad \sigma =\text{TE, TM},
\end{equation}
where
\[
D(\sigma,i\zeta,k)=1-r^2_\sigma(i\zeta,k)\,e^{4a\kappa_2(\zeta,k)},
\quad \sigma =\text{TE, TM},
\]
\begin{equation}
\label{3-11}
r_{\text{TE}}(i\zeta,k)=\frac{\kappa_1(\zeta,k)
-\kappa_2(\zeta,k)}{\kappa_1(\zeta,k)+\kappa_2(\zeta,k)}{,}\quad
r_{\text{TM}}(i\zeta,k)=\frac{\varepsilon_2(i\zeta)\kappa_1(\zeta,k)
-\varepsilon_1(i\zeta)\kappa_2(\zeta,k)}{\varepsilon_2(i\zeta)\kappa_1(\zeta,k)+
\varepsilon_1(i\zeta)\kappa_2(\zeta,k)},
\end{equation}
\[
\kappa_\alpha(\zeta,k)\equiv ik_\alpha(i\zeta,k)=
\sqrt{\varepsilon_\alpha(i\zeta)\frac{\zeta^2}{c^2}+k^2},\quad \alpha=1,2\,{.}
\]
In obtaining formula (\ref{3-10}) the integration
by parts was carried out with respect to the variable $\zeta$:
$\omega=i\zeta$. Obviously, this formula holds for
an arbitrary spectrum of electromagnetic excitations sustained
by the  configuration under consideration. Let us remember
that the symmetric configuration is considered with
nonmagnetic materials $(\varepsilon_1(\omega),
\mu_1=1|\varepsilon_2(\omega),\mu_2=1|\varepsilon_1(\omega),\mu_1=1)$.

Differentiation of the vacuum energy (\ref{3-10}) with
respect to the gap width $2a$ gives the Lifshitz
formula for the Casimir force at zero temperature~\cite{book}:
\begin{equation}
\label{3-12} F(2a)=-\frac{\partial E(2a)}{\partial
(2a)}=-\frac{\hbar}{2\pi^2}\int_0^\infty k\,dk\int_0^\infty d\zeta\,\kappa_2(\zeta,k)
\sum_\sigma \left [
r_\sigma^{-2}(i\zeta,k)e^{4a\kappa_2(\zeta,k)}-1
\right ]^{-1}.
\end{equation}
Here $\sigma= \text{TE, TM}$ as before and the same notation (\ref{3-11}) is used.

In order to reproduce  exactly the Lifshitz formula (2.9) in
\cite{Lif} for a vacuum gap  ($\varepsilon_2=\mu_2=1$) between
the identical slabs  a new
variable $p=p(k)$ should be introduced:
\begin{equation}
\label{3-13}
\frac{p\zeta}{c}\equiv\kappa_2=\left (
\frac{\zeta^2}{c^2}+k^2
\right )^{1/2}{.}
\end{equation}
Taking into account that $p(k=0)=1$ and
\[
\kappa_2k\,dk=\frac{p^2\zeta^3}{c^3}dp\,{,}
\]
we easily convert Eq.\ (\ref{3-12}) to the form
\begin{equation}
\label{3-14}
F(2a)=-\frac{\hbar^2}{2\pi^2c^3}\int_0^{\infty}\zeta^3 d\zeta \int_1^\infty p^2\,dp
\sum_\sigma \left [
r_\sigma ^{-2}(i\zeta, p)e^{4ap\zeta/c}-1
\right ]^{-1},
\end{equation}
where
\[
r^{-2}_{\text{TE}}(i\zeta,p)=\left (
\frac{\sqrt{\varepsilon_1(i\zeta)-1+p^2}+p}{\sqrt{\varepsilon_1(i\zeta)-1+p^2}-p}
\right )^2{,} \quad r^{-2}_{\text{TM}}(i\zeta,p)=\left (
\frac{\sqrt{\varepsilon_1(i\zeta)-1+p^2}+\varepsilon_1(i\zeta)
p}{\sqrt{\varepsilon_1(i\zeta)-1+p^2}-\varepsilon_1(i\zeta)p}
\right )^2{.}
\]

\section{\label{Conclusion}Conclusion}
\label{Conclusion} Our derivation of the Lifshitz formula
(\ref{3-14}) is accomplished in the framework of the
rigorous spectral summation method. We believe that in this
approach the answers are given to many questions in this
field which have needed solutions for a long time.
First of all, it is a simultaneous incorporation, on the same
footing, of the discrete and continuous parts of the
electromagnetic spectrum, with consistent treatment of  the
branch cuts in the $\omega$ plane, with consideration
of the complex eigenfrequency location and so on. Our approach
is based on a special choice of
appropriate  passes in the contour integrals which are used
for  transition to imaginary frequencies.
Contours of this type  were employed  in our preceding paper
\cite{cosmic-str} dealing with a related problem.\footnote{By
the way, it is these contours that had to be used in our
preceding paper \cite{dcyl} considering dielectric
cylinder without dispersion} The main idea of our approach
is easy to grasp. With close consideration of
Fig.~\ref{contour},  the mechanism of mutual
cancellations between contributions of the discrete and
continuous parts of the spectrum becomes evident as well as
the final result, the Lifshitz formula (\ref{3-10}). Without doubt, it is
an important advantage of our consideration.

By  making use of the contours $C_+$ and  $C_-$ shown in Fig.~\ref{contour} one
can easily `'explain'' in what way the correct Lifshitz formulae
(\ref{3-10}) and (\ref{3-11}) may be ``derived'' by summing up only
the discrete eigenfrequencies \cite{book,KNSchram,Schram,Schram-thesis,%
NPerW,Gerlach,Langbein-SSC,Langbein-JChPhys,KMM,%
Spruch-2,Spruch-3,Spruch-4}. Let us ignore the branch cuts
in the $\omega$ plane, the complex eigenfrequencies
(quasinormal modes), and the bad behavior of the left-hand
side of the frequency equation $D(\sigma,\omega)$ in the
lower half plane $\omega$. After that one may ``glue'' the
sections of the contours $C_+$  and $C_-$ which are along
the positive $\text{Re}\, \omega$ axis and drop them. As a
result, we obtain a unique contour involving the whole
$\text{Im}\,\omega$ axis and a semicircle of  large
radius in the right half plane. By using this contour in the
first integral between the square brackets in Eq.\
(\ref{3-8}) we arrive immediately at the final equation
(\ref{3-10}). Proceeding in an analogous way one may
``derive'' the Lifshitz formula in the scattering formalism
alone \cite{scatt-L,scatt-L1,scatt-L2,Mazzitelli}.
It should be noted here that the comparative analysis of
the contributions to the vacuum energy generated by
surface, waveguide, and photonic modes has recently    been carried out
by  Bordag~\cite{Bordag-11}.

In the Lifshitz formula (\ref{3-10}), written in terms of
the integral over imaginary frequencies, there are no traces
of the electromagnetic spectrum details in the problem at
hand; in other words, this formula has the same universal
form for all physically admissible material media.
Presumably, it implies that the problem of finding  the
vacuum energy with allowance for the material
characteristics of media is, on its nature, a {\it
statistical problem}. Therefore, an appropriate
mathematical method here would be the
dissipation-fluctuation theorem formulated, from the very
beginning, in terms of imaginary time. One can anticipate
that this may simplify the transition to imaginary
frequencies in the Lifshitz approach.

Many important topics are left beyond our consideration.
One of them is the allowance for the dissipation when
calculating the vacuum energy. There are many approaches to
this problem (see, for example, \cite{dissipation} and more
recent papers \cite{dissipation-r}), but until now it has been
unclear as to how to come to an agreement regarding the rigorous spectral
summation method which is based on the real
eigenfrequencies in the pertinent spectral problem.

\begin{acknowledgments}
Many topics of this paper were discussed  with  M.\
Bordag who, in particular,  drew our attention to the special class of
electromagnetic excitations in media called the wave guide waves.
We are indebted to him for fruitful collaboration.
This study was accomplished due to  financial support
from the Russian Foundation for Basic Research (Grants
No.\   10-02-01304 and No.\ 11-02-12232-ofi-m-2011), and  partial support
from the Heisenberg-Landau Program is  acknowledged.
\end{acknowledgments}



\end{document}